\documentclass[conference]{IEEEtran}
\IEEEoverridecommandlockouts
\usepackage{tabularx}
\usepackage{cite}
\usepackage{amsmath,amssymb,amsfonts}
\usepackage{algorithmic}
\usepackage{graphicx}
\usepackage{textcomp}
\usepackage{xcolor}
\usepackage{makecell}
\usepackage[skip=0.333\baselineskip]{caption}
\usepackage{subcaption}
\def\BibTeX{{\rm B\kern-.05em{\sc i\kern-.025em b}\kern-.08em
    T\kern-.1667em\lower.7ex\hbox{E}\kern-.125emX}}
\begin{document}

\title{A Dynamic Distributed Scheduler for Computing on the Edge\\
%{\footnotesize \textsuperscript{*}Note: Sub-titles are not captured in Xplore and should not be used}
\thanks{This research is supported by grants from NSF.}
}

\author{\IEEEauthorblockN{Fei Hu}
\IEEEauthorblockA{\textit{Department of Computer Science} \\
\textit{University of Colorado}\\
Boulder, CO, USA \\
Fei.Hu@colorado.edu}
\and
\IEEEauthorblockN{Kunal Mehta}
\IEEEauthorblockA{\textit{Department of Computer Science} \\
\textit{University of Colorado}\\
Boulder, CO, USA \\
Kunal.Mehta@colorado.edu}
\and
\IEEEauthorblockN{Shivakant Mishra}
\IEEEauthorblockA{\textit{Department of Computer Science} \\
\textit{University of Colorado}\\
Boulder, CO, USA \\
mishras@colorado.edu}
\and
\IEEEauthorblockN{Mohammad AlMutawa}
\IEEEauthorblockA{\textit{Computer Science Department} \\
\textit{Kuwait University}\\
Khaldiya, Kuwait \\
almutawa@cs.ku.edu.kw}
}
\maketitle

\begin{abstract}
Edge computing has become a promising computing paradigm for building IoT (Internet of Things) applications, particularly for applications with specific constraints such as latency or privacy requirements. Due to resource constraints at the edge, it is important to efficiently utilize all available computing resources to satisfy these constraints. A key challenge in utilizing these computing resources is the scheduling of different computing tasks in a dynamically varying, highly hybrid computing environment. This paper described the design, implementation, and evaluation of a  distributed scheduler for the edge that constantly monitors the current state of the computing infrastructure and dynamically schedules various computing tasks to ensure that all application constraints are met. This scheduler has been extensively evaluated with real-world AI applications under different scenarios and demonstrates that it outperforms current scheduling approaches in satisfying various application constraints.
\end{abstract}


\section{Introduction}
In recent years, Edge AI systems \cite{edgeAI1, edgeAI2} have emerged as a promising solution for handling the growing demand for AI workloads by bringing computation closer to the data sources and reducing latency \cite{edgecomputing1, edgecomputing2}. However, to fully realize the potential of these systems, efficient utilization of computing resources is critical. One of the key factors that affect the performance of Edge AI systems is the scheduling algorithm used to assign tasks to computing nodes.

This paper proposes a new distributed scheduling method that addresses this challenge by leveraging the computing capacities of all nodes, dynamically adjusting the scheduling decisions based on the system's current state, and taking various constraints into account. These constraints include real-time requirements, computation capability, network bandwidth, and other factors that impact the performance of the system. By considering a range of objectives and constraints, the proposed algorithm offers a promising approach to optimize the deployment of AI workloads in real-world scenarios.

The contribution of this research is the development of more efficient and effective distributed scheduling algorithms for the edge platform, with potential applications in various domains such as smart factories, healthcare, and autonomous vehicles. The proposed algorithm has the potential to enhance the performance of Edge AI systems while minimizing latency and optimizing the utilization of computing resources. Additionally, this paper provides a comprehensive evaluation of the proposed method's performance, demonstrating its effectiveness in real-world scenarios.

The remainder of this paper is organized as follows: Section II reviews related work in distributed scheduling algorithms for the edge platform. Section III presents the proposed scheduling algorithm in detail, highlighting its advantages over existing methods. Section IV describes the evaluation methodology and presents the results of the experiments conducted to validate the proposed algorithm's effectiveness. Section V discusses the implications of the proposed algorithm and outlines future research directions. Finally, Section VI concludes the paper by summarizing the main contributions and highlighting their significance for the development of Edge AI systems.

 \section{Related Work}
End devices are capable of handling some tasks on their own with the available computation resources. The scheduling problem at the edge involves distributing tasks to these nodes. However, there are several challenges currently faced in scheduling, including the following aspects:

Incomplete Information: Due to privacy and security concerns, some devices may be unwilling to share all their information with other devices. Additionally, network delays may prevent information synchronization from occurring in a timely manner \cite{incompleteinfo1, incompleteinfo2 }.

Constraint Cooperativeness: Incomplete cooperation between end devices can limit their ability to work together, as some devices may not always be willing to share their resources with others.

Task and Trust Constraints: Some users may submit tasks only to specific nodes for privacy or security reasons, and these tasks cannot be shared with any other devices. Furthermore, certain nodes may only accept specific requests; for example, node A may only process images, while node B may only handle video \cite{trustconstraint1, trustconstraint2}.

Dynamic Environment: The edge network is not always stable, as the workload may vary depending on the task, and the physical location of mobile devices may change. The information needs to be updated with nearby nodes, which could be a scheduler manager or other devices to cooperate with \cite{dynamicenvironment1, dynamicenvironment2}.

To address these challenges, current research in edge computing focuses on meeting objectives such as low latency, low energy consumption, and multi-objectives. These efforts primarily fall into two categories. The first is a centralized solution where a scheduler manager offloads tasks to edge devices with limited memory, battery, and computational power. However, this approach comes with the heavy overhead of the manager and large transfer and bandwidth costs associated with pushing data and computation tasks to end nodes often results in large transfer and bandwidth costs. 

For example, Zhu et al. proposed a Mixed Integer Linear Programming based approach in their paper \cite{zhu2013optimization} to minimize latency. Zhou et al. proposed a mixed-criticality task model in their paper \cite{zhou2015thermal} to process as many latency-critical tasks as possible while also processing many low-criticality tasks. However, in real scenarios, many end nodes may refuse to share all information with the manager for privacy or security concerns, making the centralized scheduler approach fail.

The second category is a distributed solution where multiple nodes coordinate to schedule tasks. Numerous solutions and protocols have been designed in this direction. For instance, Ahmad et al. proposed a game theory-based solution in their paper \cite{ahmad2008using} to optimize the solution and save energy. Bertuccelli et al. designed a decentralized task allocation protocol under dynamic environments in their paper \cite{bertuccelli2009real}. Zheng et al. presented an integrated formulation for optimizing task placement in distributed systems in their paper \cite{zheng2007definition}. This method considers task priority to meet end-to-end deadline constraints and minimize total latencies.

Due to limited computing capacity, end devices are usually unable to process user requests by themselves. Paper \cite{chang2015heterogeneous, khelifi2018bringing, chen2020exploring} discuss how end devices can extend their processing scope and complete user requests by collaborating with other access nodes.

However, all of that research is based on mathematical modeling, which is unable to fully account for the complexity and variability of dynamic edge networks. Therefore, we propose a brand-new solution: a dynamic distributed scheduling algorithm based on real-world evaluation, to support computing at the edge.

\section{Edge AI Architecture}

To date, almost all relevant research on scheduling tasks at the edge has relied on mathematical modeling and theoretical analysis. One limitation of this approach is its limited practical applicability due to the need to consider multiple variables in real-world scenarios. Consequently, our scheduler is based on evaluation results that reflect the computation capacity of different devices and the corresponding end-to-end latency for specific applications.

\subsection{Scheduler Design Principles}

An IoT environment at the edge is typically highly hybrid and dynamic in nature. Computing devices range from low-end, resource-constrained devices to moderately powerful edge servers. Further, the communication and compute load can vary significantly over time depending on what events take place at the edge. So, it is important to ensure that the scheduler takes into account these dynamically varying factors while deciding task scheduling. Consequently, a key challenge in building a distributed scheduler for a dynamic edge environment is how to balance the overhead incurred by maintaining current up-to-date information about the computing environment and scheduling decision-making with the potential gains in application performance.

Since a centralized scheduler cannot support non-cooperative end devices, our system will use a distributed architecture consisting of {\it two levels}. Each node, including edge servers, end devices, and the cloud, runs a scheduler component, with the edge server scheduler acting as the coordinator. At the lower level, each device records its current computing status (CPU load, network latency to the edge server, remaining battery power (if applicable), etc.) periodically, and shares its status with the edge server. 
The edge server maintains the current status of each device based on this information and uses it to make distributed scheduling decisions. 

Since the network and CPU load conditions can change quickly and it takes time to gather the current state of the system, it is realistic to assume that the scheduler will have to make a scheduling decision based on system conditions that may be slightly out of date. Based on this observation, our primary guiding principle is to minimize communication among devices at runtime when making scheduling decisions, i.e. {\it schedule a task locally if current conditions indicate that all constraints can be met by computing locally}. If the local node has the capability to address the current task while considering constraints, it will process the task locally; otherwise, the job will be sent to the edge server, which then may offload it to other end nodes depending on the current system conditions. Since the edge server acts as the coordinator, we assume that it always makes the best global choice.

\subsection{Device Profiling}

Edge devices are usually optimized for specific types of applications, such as face detection and gesture detection. We approximate the relationship between system performance and decision-making overhead by evaluating one application to represent one type of application.

Based on the above evaluation results, we assume that each device knows its own capabilities. For example, it knows how much time it will take to perform face detection on a specific image. The edge server component is aware of the capabilities of all devices, and end devices regularly update their profiles. This updating information includes current device loads, network bandwidth, and so on, given the dynamic nature of the system.

For task x, the processing time on node e can be formulated as $T_{task}(x, e) = T_{trans}(x,e) + T_{que}(x,e) + T_{process}(x,e) + T_{re}(x,es)$
The variables $T_{trans}$, $T_{que}$, $T_{process}$, and $T_{re}$ represent the time taken for task x to be transmitted to the device e, queued, processed, and received, respectively. Among these, the transmitting and receiving times depend on the data size and bandwidth. To simulate a practical scenario where some requests may not be received successfully, we use UDP to send the requests.

\subsection{Architecture Components: Upper Level}
\begin{figure}[ht]
    \centering
\includegraphics[width=0.45\textwidth]{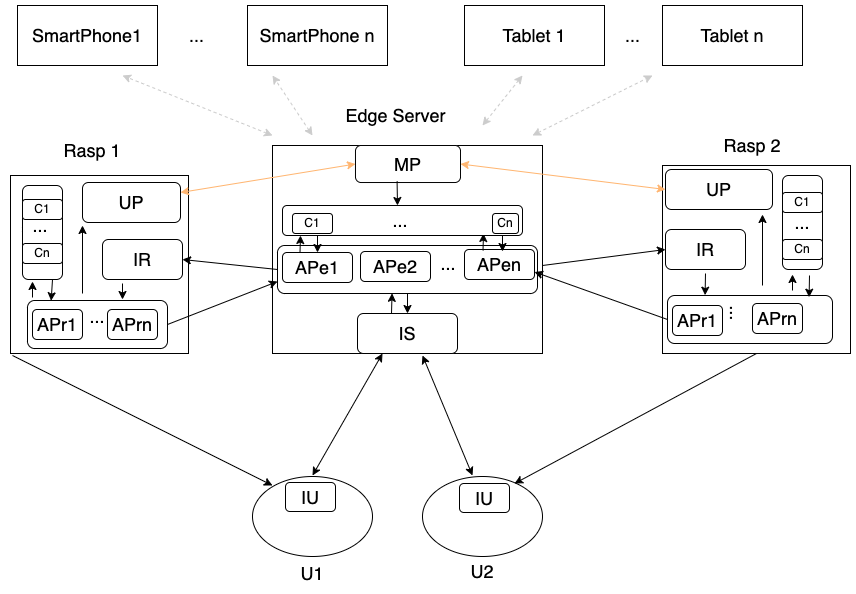}
    \caption{Architecture}
    \label{fig: architecture}
\end{figure}

The proposed architecture, as depicted in Figure \ref{fig: architecture}, is designed to support Edge AI applications by distributing tasks to end devices based on their relative locations, ensuring that the data is acquired quickly and efficiently. When a user initiates an application request, the edge server identifies the nearby end devices and assigns the task to them. For instance, in a crowded mall scenario, the edge server will stimulate end devices that are in close proximity to the user, such as cameras or video modules, to obtain a stream of the crowd and detect the target person with higher accuracy.
This architecture comprises three major components, namely the edge server, end devices, and mobile users. 
%In this section, we will provide a detailed overview of each component and describe the workflow of the system. 

\subsubsection{Edge Server}

The edge server, which acts as the central coordinator, consists of four key components: Interface Server (IS), edge server application pools (APe1 to APen), container pool, and Maintain Profile (MP) module. The IS receives user requests, analyses them, and then transfers them to the appropriate APe. The APe contains all the applications that the system supports, such as face detection, object detection, and other augmented reality applications. Depending on the requested information, the APe decides whether to process the task locally or offload it to other end devices. If the latter, the APe connects with the end device Interface Server (e.g. Interface Raspberry pi (IR) on Rasp 1 and Rasp 2), which forwards the requests to the corresponding interface server. Additionally, the APe receives the results from the end devices through the end device application, such as APr on Raspberry Pis. On the other hand, if the edge server decides to process the application locally, it invokes the appropriate containers from the container pool and waits for the results upon completion. The MP module is an essential component that supports the dynamic distributed scheduling algorithm. It connects with other Update Profile (UP) modules to collect profile information of all other end devices and maintain a global profile table. The scheduler uses this table to adjust the distribution of tasks to meet user requests and efficiently utilize end computing resources in real time.

\subsubsection{End Devices}
Any device that has computing resources, such as Raspberry Pis, SmartPhones, and Tablets, can serve as an end device within the architecture. Before joining the system, the device needs to be certified to ensure it meets the requirements. Each end device consists of four components, as illustrated in Rasp 1. The Update Profile (UP) component is responsible for updating the device's profile information, such as the device ID and CPU load, with the edge server. The Interface Receiver (IR) component receives requests from the edge server, and the Application pool (APr) contains applications that the device supports. Unlike the edge server, which can handle a variety of applications, end devices typically support specific applications due to their limited computing resources. The Containers pool provides the runtime environment defined by APr for the applications to run on the end device.

\subsubsection{Mobile Users}

In our research project, we have developed two Android applications that communicate with each other using socket programming on the network. The devices use one socket on both the client and server sides to send and receive messages. Our server can accept multiple clients and communicate with them. The server listens at a specific port, and clients can connect to it after establishing a connection. The server runs on a separate thread in our application.

\begin{figure}[ht]
    \centering
\includegraphics[width=0.3\textwidth]{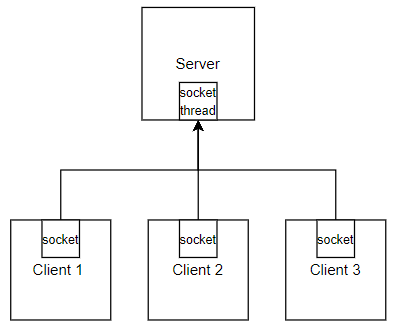}
    \caption{Client server socket communication}
    \label{fig: Socket communication}
\end{figure}

The client interface includes two buttons and three text fields for input. When the client presses the "Connect" button, it starts a new thread that attempts to connect to the server specified by the IP address and port number. The "Send" button sends the entered message to the server, and incoming and outgoing messages are displayed on the screen in different colors for better readability.

Clients can send their location, constraints, and the desired function to the server, which then sends the data to the nearest edge server based on the given constraints. If the constraints are met, the results are sent back to the client. We create a separate thread to run our server, which accepts incoming connections.

Using socket communication has several benefits over RESTful methods, as it does not rely on external environments. Socket programming uses the TCP protocol, allowing for communication on various devices such as mobile phones, Raspberry Pis, and Arduinos.

\subsection {Architecture Workflow}

Figure~\ref{fig: architecture} depicts the architecture of a distributed scheduler system. The system receives a request from a user through Interface User(IU) that includes the requested application ID, user location, and other relevant information. The request is sent to the Information Server (IS), which forwards it to the corresponding Edge Server Applications (APe), such as a face detection application.

Based on the user's location, the APe determines which Raspberry Pi device to take pictures of and sends the request to the Interface Raspberry Pi (IR) on that device. The APe also sends a reply to the user.

The IR receives the request from the edge server and sends it to the raspberry pi Application (APr). The APr sends a request to the camera and receives images from it. For each image, the APr makes scheduling decisions based on the current load. It gets load information from the Update Profile module (UP), which collects and updates a table reflecting the current load. If the end device has the capacity to run the image processing locally, then the APr sends the image to the local container. If not, the image is sent to the APe on the edge server.

The APe receives images from the APr and makes decisions to send them to another end device or run them locally based on a scheduling strategy. If it runs locally, it starts the container; otherwise, it sends the image to the APr on another device. The APe and APr distinguish among different requests through different byte types. For example, if the APe receives a request from the IS, it should contain the Application ID, and then the system enters the function that makes the decision to send the request to a closed-end device based on location. If the request comes with an image, the APe knows it is for an image processing request.

The UP keeps track of the current load and updates it with the Management Processor (MP) regularly. The APe gets this data through shared memory when making decisions. This architecture enables distributed scheduling of image processing tasks in a scalable and efficient manner.

\section{Device Profile Evaluation}
To meet our latency requirements, we need to determine the processing time for each task on devices with variable loads. Since containers are lightweight and are a suitable platform for deploying applications on edge networks \cite{container1, container2}, we conducted a test on the container to observe the impact of load on its performance.
 
\subsection{Face Detection Container}
\begin{figure}[hbt!]
    \centering
\begin{subfigure}{.46\linewidth}
  \includegraphics[width=\linewidth]{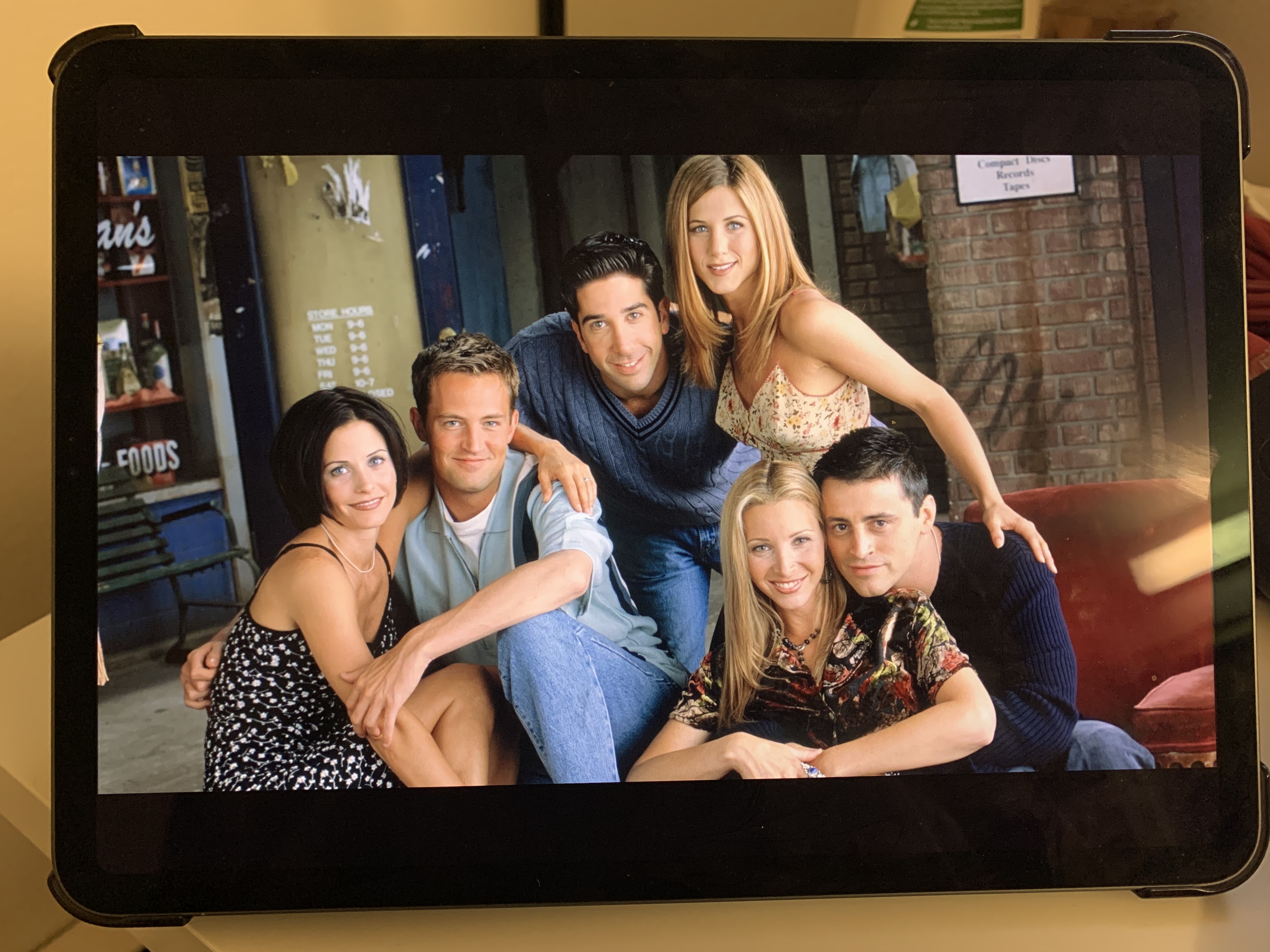}
  \caption{Original Picture}
  \label{fig:origin_pictures}
\end{subfigure}
\hfill % <-- "\hfill"
\begin{subfigure}{.48\linewidth}
    \centering
   \includegraphics[width=\linewidth]{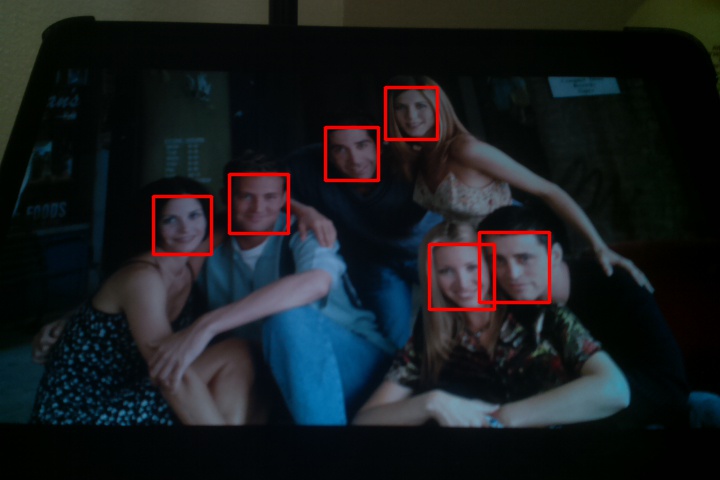}
  \caption{Face Detection Result}
  \label{fig:face_detection}
\end{subfigure}

\caption{An Example of Face Detection}
\label{fig:face_detection_example}
\end{figure}

Face detection is a computer vision task that involves locating and detecting human faces in images or videos \cite{facedetection1}. There are several algorithms available for face detection, including the Viola-Jones algorithm, Histogram of Oriented Gradients (HOG) algorithm, Convolutional Neural Networks (CNN) algorithm, and Skin color detection algorithm. In this project, we use the Viola-Jones algorithm for face detection. Compared to other algorithms, the Viola-Jones algorithm offers several advantages, including its relatively fast and real-time performance, robustness to variations in lighting and image quality, accurate detection, versatility, and ease of implementation. For this face detection application, we created a corresponding image with 18 layers, which included a basic Ubuntu operating system and necessary system installations. The docker container image size is 1.08GB. When a picture with people is processed, the container detects all the faces in the image and highlights them with circles, as shown in Figure~\ref{fig:face_detection_example}.

The runtime of the face detection algorithm can be influenced by several factors, including hardware. Faster processors and more memory can speed up the runtime of the algorithm. To demonstrate the runtime difference, we will use the same algorithm and image on two different devices, whose systems specifications are shown in Table \ref{table:sysconfig}  The size of the input image being processed is another factor that can influence the runtime of the face detection algorithm. The larger the input image, the longer it will take to process, as the algorithm needs to examine more pixels to detect faces, which requires more computation. Table \ref{tab1} shows the runtime of the above-mentioned face detection container for different image sizes on the edge server.

%\vspace{-0.1in}
\begin{table}[htbp]
\caption{System Configuration}
\vspace{-0.1in}
\begin{center}
\resizebox{0.45\textwidth}{!}{%
\begin{tabular}{|c|c|}
 \hline
 \textbf{Component Name} & \textbf{Specifications} \\
  \hline
 Edge Server  & \thead{2.3GHz Dual-Core Intel Core i5,\\ 8 GB RAM,  256GB Disk} \\
  \hline
 Raspberry Pi &  \thead{Quad core Cortex-A72 (ARM v8) 64-bit \\ 8GB RAM, 1.8GHz Clock Speed} \\
 \hline
 Smart Phone &  \thead{Octa-core (4x2.3 GHz Mongoose,\\ 4x1.6 GHz Cortex-A53), 4GB RAM} \\
 \hline
\end{tabular}%
}
\label{table:sysconfig}
\end{center}
%\vspace{-0.25in}
\end{table}

\begin{table}[htbp]
\caption{Runtime for Different Image Size}
\begin{center}
\begin{tabular}{|c|c|c|c|c|c|}
\hline
Image Size (KB) & 29 & 87 & 133 & 172 & 259 \\
\hline
Runtime (ms) & 223 & 417 & 615 & 798 & 1163 \\
\hline
\end{tabular}
\label{tab1}
\end{center}
\end{table}

\subsection{Profile Evaluation Scenarios}

In this paper, we present an experimental study to evaluate the processing time of containers in different scenarios under varying workloads. Our study is motivated by the need to design an efficient and dynamic distributed scheduler that can allocate tasks to containers based on the current load and the impact of the new task on the existing ones.

\begin{table*}[htbp]
\caption{Profile of Cold Container on the Edge Server}
\begin{center}
\resizebox{0.7\textwidth}{!}{%
\begin{tabular}{|c|c|c|c|c|c|}
\hline
Container Number & 1 & 3 & 5 & 8 & 11  \\
\hline
Run Time of Existing Container (ms) & 63887 & 121766 & 226044 & 328269 & 716767  \\
\hline
Run Time of the New Container (ms) & 52554 & 71788 & 106596 & 165717 & 437846 \\
\hline
\end{tabular}%
}
\label{tab: cold start on edge server}
\end{center}
\end{table*}

\begin{table*}[htbp]
\caption{Profile of Cold Container on the Raspberry Pi}
\begin{center}
\resizebox{0.8\textwidth}{!}{%
\begin{tabular}{|c|c|c|c|c|c|c|}
\hline
Container Number & 1 & 2 & 3 & 4 & 5 & 6  \\
\hline
Processing time of existing container (ms) & 160802 & 198529 & 248812 & 313466 & 424130 & 520442 \\
\hline
Processing time of the new container (ms) & 168279 & 179280 & 188633 & 211136 & 241222 & 249413 \\
\hline
\end{tabular}%
}
\label{tab: cold start for Rasp}
\end{center}
\end{table*}

Based on the status of existing containers and the new container, there are four scenarios to represent the processing time of containers. In the first scenario, we started n containers and waited for them to warm up before sending a face detection request to all of them simultaneously. We recorded the start time (S) and the end time (E) of each container to calculate its processing time, where E-S is the time taken to process the request. We repeated this experiment 10 times for each situation.

\begin{table*}[htbp]
\caption{Profile of Warm Container on the Edge Server}
\begin{center}
\resizebox{0.65\textwidth}{!}{%
\begin{tabular}{|c|c|c|c|c|c|c|c|c|}
\hline
Container Number & 1 & 2 & 3 & 4 & 5 & 6 & 7 & 8 \\
\hline
Average Time (ms) & 223 & 273 & 366 & 464 & 540 & 644 & 837 & 947 \\
\hline
Total Time (ms) & 11193 & 6930 & 6216 & 5951 & 5794 & 5507 & 6020 & 6099 \\
\hline
\end{tabular}%
}
\label{tab: edge server profile}
\end{center}
\end{table*}

\begin{table*}[htbp]
\caption{Profile of Warm Container on the Raspberry Pi}
\begin{center}
\resizebox{0.58\textwidth}{!}{%
\begin{tabular}{|c|c|c|c|c|c|c|}
\hline
Container Number & 1 & 2 & 2 & 4 & 5 & 6 \\
\hline
Average Time (ms) & 597 & 613 & 651 & 860 & 1071 & 1290 \\
\hline
Total Time (ms) & 29934 & 15399 & 11072 & 11042 & 11043 & 11074  \\
\hline
\end{tabular}%
}
\label{tab: Rasp profile}
\end{center}
\end{table*}

In the second scenario, we recorded the start time (S) before starting to create n cold containers. We then calculated the processing time of each container by recording the start time ($S_i$), running the application, and recording the end time ($E_i$). We also calculated the total processing time by adding the time taken to create the containers to the processing time of the last container.

The third and fourth scenarios were designed to simulate real-world situations where an additional job needs to be allocated to n existing containers. In scenario three, we started n containers and an additional container simultaneously. When the n containers started processing the face detection request, we sent a request to the additional container. We recorded the processing time of each container and the additional container to analyze the impact of the new job on the existing ones. In scenario four, we repeated the same process as scenario three but created the additional container only when it received the request.

Our experiments provide insights into the performance of containers under varying workloads and highlight the need for a dynamic distributed scheduler. By analyzing the processing times of containers under different scenarios, we propose a new distributed scheduling method that is based on the real-world evaluation. The proposed method is designed to optimize the deployment of containers while minimizing latency and utilizing computing resources efficiently.

\subsection{Profile Evaluation Results}

Table \ref{tab: cold start on edge server} and Table \ref{tab: cold start for Rasp} show the test results for cold start containers on the edge server and Raspberry Pi respectively. The row for "Run Time of Existing Container" corresponds to Scenario 2, while the row for "Run Time of New Container" corresponds to Scenario 4, as previously mentioned. As the number of containers increases, the time required to start a cold container also increases, along with the time required to start the other n-1 containers. Similarly, the more containers that need to start, the longer it takes for an extra cold container to start.

Since starting a cold container takes a significant amount of time (e.g., starting an extra container takes 52554ms with just one), it is not practical to start a cold container upon receiving a request.

Table \ref{tab: edge server profile} and Table \ref{tab: Rasp profile} show the test results for cold start containers on the edge server and Raspberry Pi respectively. show the test results for cold start containers on the edge server and Raspberry Pi, respectively. The row labeled "Average Time" represents the average processing time of one image in a container, which corresponds to Scenario 2 and Scenario 4. Obtaining the run time of an extra container with n warm containers is similar to obtaining the average processing time for n+1 containers.

Both tables show that as the number of containers increases, the average processing time also increases. This is expected, as on the four-core edge server, more containers will compete for the limited computing resources, which decreases the processing time for each task.

As shown in Table \ref{tab: edge server profile}, increasing the number of containers from 1 to 2 decreases the total processing time of 50 images from 11193ms to 6930ms. This is because with one container running, one core is occupied and the idle CPU is enough, running 2 containers could increase the utility of the CPU and achieve about half of the total processing time. Based on our observations, running 3 containers would occupy about 75\% of the system CPU, and the scheduling program occupies about 15\% of the user CPU, leaving about 10\% of idle CPU. With 4 containers running, the idle CPU gets close to 0\%. Therefore, the total time shows a slight decrease from 6216ms to 5951ms. After that, as the CPU load is almost fully occupied, increasing the number of containers could not improve performance anymore, and the total time stabilizes at around 6000ms.

\section{Evaluation and Results Analysis}

\subsection{Application}

\begin{figure}[ht]
    \centering
\includegraphics[width=0.45\textwidth]{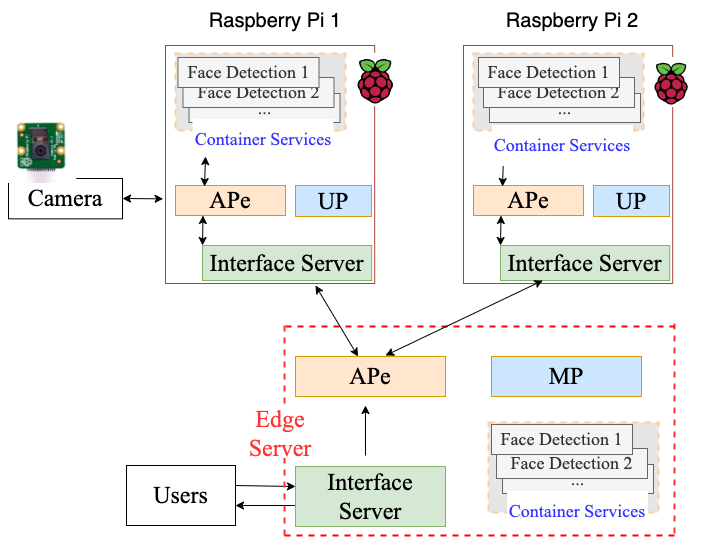}
    \caption{Experiment Application}
    \label{fig: application}
\end{figure}

The experimental application is illustrated in Figure \ref{fig: application}. It provides a visual representation of the components of the architecture and their interactions. The architecture includes an edge server, two Raspberry Pis, a mobile phone, and a camera. The system specifications are listed in Table \ref{table:sysconfig}.

\subsubsection{Application Components}
The edge server comprises an interface server, an APe, a container pool, and an MP. The interface server is responsible for communicating with users and analyzing their requests. The APe is a multi-threaded program that manages the local container pool and interacts with other end devices. In our simulation, the APe corresponds to the face detection application, and the container pool includes only face detection containers. The MP continuously receives profile information and updates the profile table.

\begin{figure*}[hbt!]
    \centering
\begin{subfigure}{.46\textwidth}
 \centering
  \includegraphics[width=\linewidth]{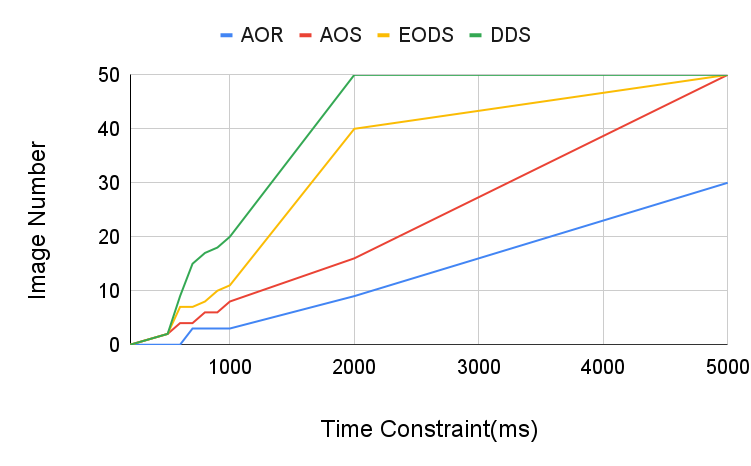}
  \caption{Time Interval (50ms)}
  \label{fig:50_50}
\end{subfigure}
\hfill % <-- "\hfill"
\begin{subfigure}{.46\textwidth}
\centering
   \includegraphics[width=\linewidth]{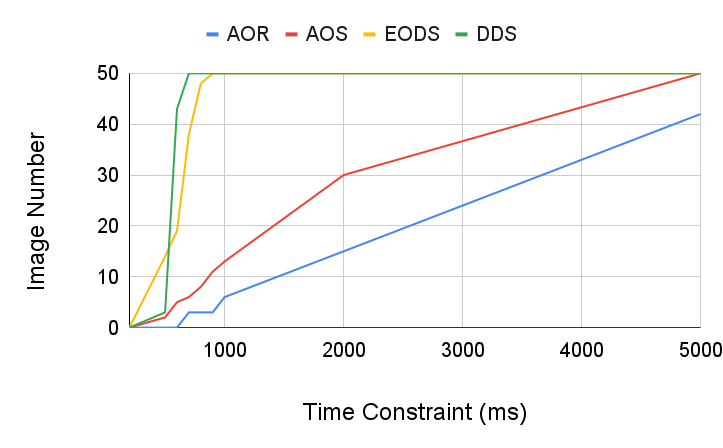}

  \caption{Time Interval (100ms)}
  \label{fig:50_100}
\end{subfigure}

\medskip % create some *vertical* separation between the graphs
\begin{subfigure}{.46\textwidth}
\centering
  \includegraphics[width=\linewidth]{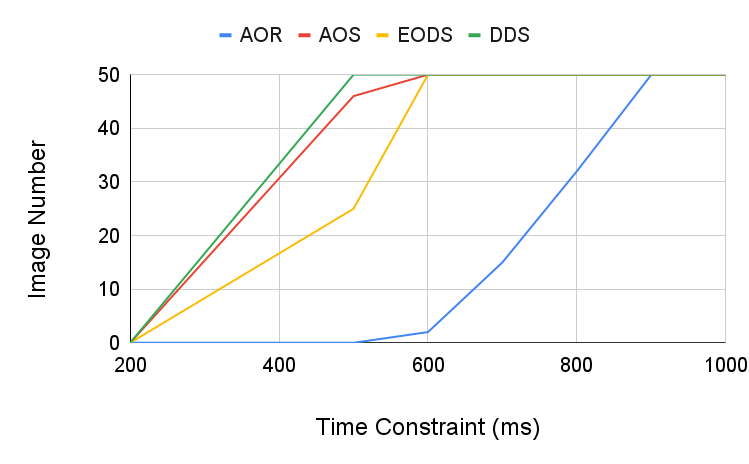}
  \caption{Time Interval (200ms)}
  \label{fig:50_200}
\end{subfigure}
\hfill % <-- "\hfill"
\begin{subfigure}{.46\textwidth}
\centering
  \includegraphics[width=\linewidth]{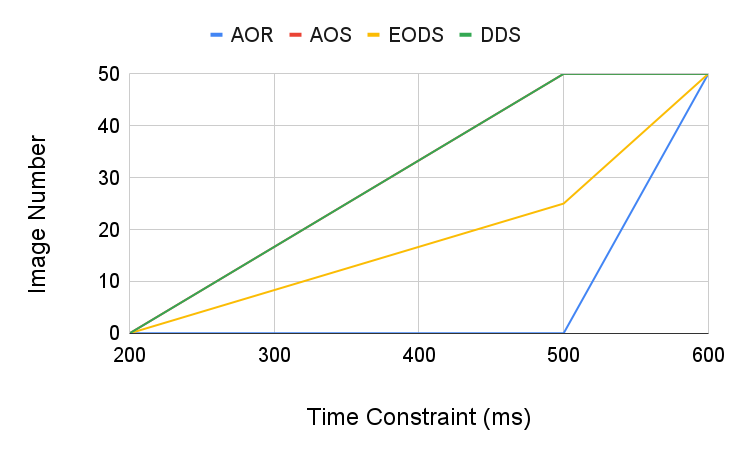}
  \caption{Time Interval (500ms)}
  \label{fig:50_500}
\end{subfigure}

\caption{50 Images}
\label{fig:50}
\end{figure*}

In this architecture, we have incorporated two Raspberry Pis, each with an interface server, an APr, a container pool, and a UP. The interface server receives requests from the edge server and analyzes request information. The APr is a multi-threaded program that manages the local container pool and interacts with sensors connected to it. Similar to the edge server, the APr on each Raspberry Pi corresponds to a face detection application, and the container pool includes only face detection containers.  As shown in Figure \ref{fig: application}, a camera is connected to Raspberry Pi 1, which captures images for the face detection application.

\subsubsection{Application Workflow}
This architecture functions as follows: During the initial stage, two Raspberry Pi devices request to join the distributed architecture. Upon approval by the edge server, they establish a connection with it and update their profile information through the UP module.

When a user sends a request containing information such as the application type, user IP address, and location, the edge server interface receives it and sends it to the APe module. APe then forwards the request to Raspberry Pi 1, and updates which node the user's request will be directed to. The APr module activates the camera to capture images and sends them back to APe. APr consists of three main threads: one thread receives images from the camera and stores them in an original image queue; another thread makes a decision for the top image in the queue based on scheduling rules, determining whether to process it locally or send it to the edge server. If the predicted processing time, based on profile evaluation, is within the time constraint, the task is processed locally. For example, if a job's running time is 597ms without other running containers, and the time constraint is less than this number, the task is sent to the edge server. Conversely, if the time constraint is greater than 597ms, the current task is processed locally on the container. For images to be processed locally, if available containers exist, the image is sent to one of them. Otherwise, the image number is saved in a separate queue called $q_image$ for later processing. As the available container ports are stored in the q queue, the number of available containers can be determined by the size of q. The third thread is responsible for receiving replies from containers. When feedback is received from a container indicating that it is available for the next image, the thread also checks whether there are images waiting to be processed in the $q_image$ queue. If images are present, the container continues processing with the next image. Otherwise, the container number is pushed back to the q queue.

The edge server APe comprises three threads. The first thread receives images from the end device and stores them in an original queue. The second thread receives replies from containers, similar to Raspberry Pi 1. The third thread makes decisions for all images sent to the edge server. To make full use of computing resources on end devices, the main scheduler predicts the processing time on Raspberry Pi 2 based on profile information. Shared Memory is used to speed up reading speed. If the task can be processed on Raspberry Pi 2, it is transformed and processed there. Otherwise, the task is processed locally. Similar to the solution on Raspberry Pi 1, the current image is sent to available containers or saved in the $q_image$ queue waiting for available containers.

The Raspberry Pi 2 APr module includes three threads. One thread is responsible for receiving images, another for receiving feedback from local containers (as in Raspberry Pi 1), and the third thread is for making decisions. For all jobs sent to it, the third thread processes them locally. Raspberry Pi 2 UP updates its profile information, including the number of running containers, every 20ms.

\begin{figure*}[hbt!]
    \centering
\begin{subfigure}{.46\linewidth}
  \includegraphics[width=\linewidth]{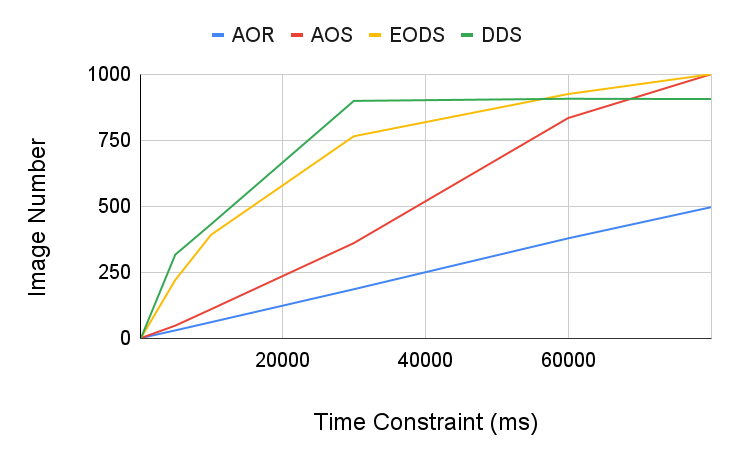}
  \caption{Time Interval (50ms)}
  \label{fig:1000_50}
\end{subfigure}
\hfill % <-- "\hfill"
\begin{subfigure}{.46\linewidth}
    \centering
   \includegraphics[width=\linewidth]{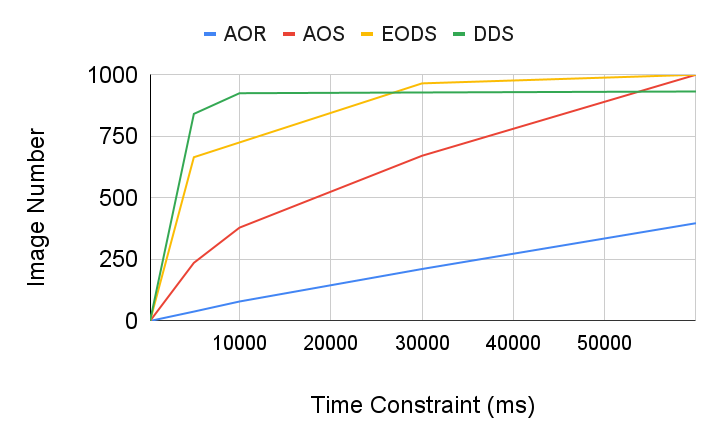}
  \caption{Time Interval (100ms)}
  \label{fig:1000_100}
\end{subfigure}

\caption{1000 Images}
\label{fig:1000}
\end{figure*}

\subsection{Results and Analysis}

In order to compare our Dynamic Distributed Scheduler (DDS) with other approaches, we created three comparison groups. The first group is to run ALL On the Raspberry Pi (AOR), without utilizing any edge server computing resources. The second comparison group involved transmitting all images to the edge server and Run All on the Edge Sever (AOE). We created a third comparison group where the Raspberry Pi was responsible for processing images with odd-numbered sequences, while all images with even-numbered sequences were transmitted to the edge server for processing, named Even Odd Distributed Scheduling (EODS).

\begin{figure}[ht]
    \centering
\includegraphics[width=0.80\linewidth]{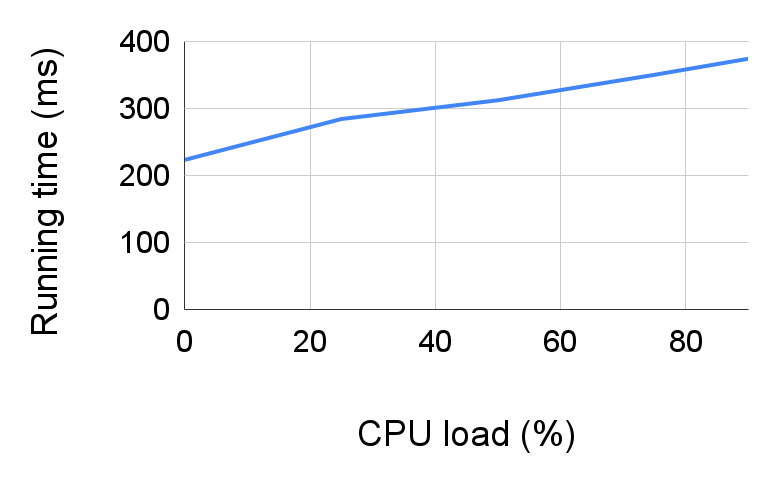}
    \caption{Relationship between CPU Load and performance}
    \label{fig:cpuload}
\end{figure}

\subsubsection{Set 50 images}
This experiment explores the performance of four different scheduling algorithms when an end device continuously receives a stream of images from a buffer module after receiving an application request. The evaluation results of testing 50 images with each scheduling algorithm are presented in Figure~\ref{fig:50}.

As the time interval between two images can significantly impact the performance of the system, the experiment sets it as a parameter and tests the algorithms' performance at four different time intervals: 50ms, 100ms, 200ms, and 500ms. The evaluation results for each time interval are presented in Figure~\ref{fig:50_50}, Figure~\ref{fig:50_100}, Figure~\ref{fig:50_200}, and Figure~\ref{fig:50_500}, respectively. Each subfigure in the evaluation results reflects the relationship between the number of images that meet the requirements and time constraints.

As shown in Figure \ref{fig:50}, we can observe that with an increase in time constraints, a greater number of images meet the time requirements for all scheduling algorithms. For instance, in Figure \ref{fig:50_50}, when the time constraint is set to 500ms, none of the images meet the constraint while running all images on Raspberry Pi. However, when the time constraint is increased to 5000ms, 30 out of the 50 images meet the requirement. Furthermore, we can observe that as the time interval increases, more images are able to meet the requirements. For example, when comparing the results of running all images on Raspberry Pi with a time constraint of 1000ms in Figure \ref{fig:50_50} and Figure \ref{fig:50_500}, only 3 images meet the requirements when the time interval is set to 50ms, but when the time interval increases to 500ms, all images are able to meet the requirements. This trend is reasonable because a longer time interval leads to a shorter queuing time and fewer simultaneous running containers, which reduces the computing load and decreases the processing time for each application as images are processed in a more efficient manner.

\begin{figure*}[hbt!]
    \centering
\begin{subfigure}{.46\linewidth}
  \includegraphics[width=\linewidth]{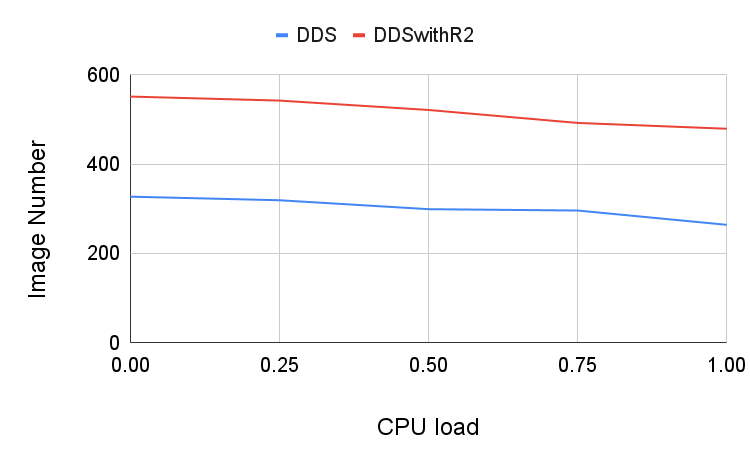}
  \caption{Time Constraint (5000ms)}
  \label{fig:5000_DDS_vs_DDSwithR2}
\end{subfigure}
\hfill % <-- "\hfill"
\begin{subfigure}{.46\linewidth}
    \centering
   \includegraphics[width=\linewidth]{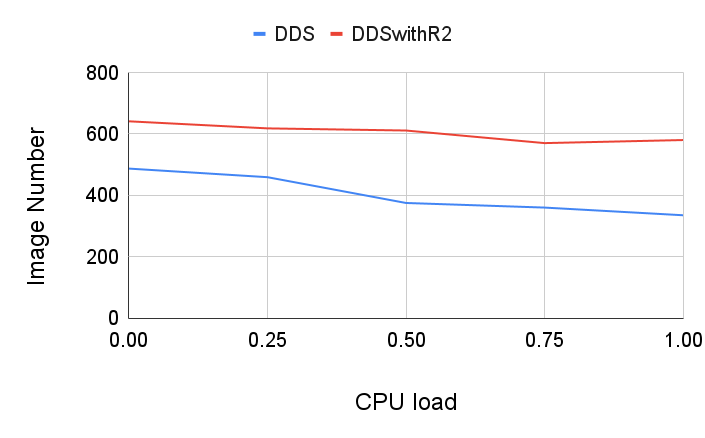}
  \caption{Time Constraint (10000ms)}
  \label{fig:10000_DDS_vs_DDSwithR2}
\end{subfigure}

\caption{DDS vs DDSwithR2}
\label{fig:DDS_vs_DDSwithR2}
\end{figure*}

From Figure~\ref{fig:50}, insightful conclusions can be drawn regarding scheduling algorithm design and building scheduling systems.
\begin{itemize}

\item It is important to set the minimum time constraint required for all requests. If the time constraint is too short, none of the scheduling algorithms can improve performance. For instance, when the time constraint is less than 200ms, none of the four scheduling algorithms meet the image processing requirements. Therefore, any application requests with a time constraint less than this time should be rejected. 

\item Computing nodes with powerful resources outperform weaker ones, as observed in all subfigures of Figure~\ref{fig:50}, where the edge server always performs better than the end device. This trend is also observed in the device profile evaluation section. 

\item For requests with loose constraints such as long time intervals and long time constraints, running everything on the edge server has the best output and the least overhead. 

\item Distributed scheduling systems such as Even Odd distributed scheduling and the Dynamic Distributed Scheduler perform better than running all images on either the edge server or the end device. This observation is supported by all the cases in the experiment. 

\item The Dynamic Distributed Scheduler is better than the Even Odd Distributed Scheduler, except when the edge server is heavily loaded. For example, in Figure~\ref{fig:50_50}, when the time constraint is 500ms, the Raspberry Pi is unable to process any image within the time limit, leading to a heavy load on the edge server and a long queuing list. In this case, the Dynamic Distributed Scheduler is similar to running everything on the edge server.

\end{itemize}

\subsubsection{Set 1000 images}
From the experiments with 50 images, we show the priority of distributed scheduling mechanism and dynamic scheduling is better than static scheduling like Even Odd Distributed Scheduling. And we further explore DDS's performance with continued streaming frames. As we know when the time interval is relatively long, the AOS scheduling algorithm performs well, so in this experiment, we set the buffer module to send 1000 images with a time interval between images 50ms and 100ms respectively.

Figure \ref{fig:1000} displays a comparative analysis of the performance of four scheduling algorithms with regard to the number of images meeting the time constraint over a range of time constraints. The x-axis of the graph represents the time constraint with values ranging from 200ms to 80,000ms, whereas the y-axis represents the number of images (out of 1000 images) that meet the time constraint. The evaluation results for a time interval of 50ms and 100ms are shown in Figure \ref{fig:1000_50} and Figure \ref{fig:1000_100}, respectively.

A comparison of the DDS algorithm with the other algorithms,  indicates the following observations: 

\begin{itemize}
\item DDS performs better when the time constraint is less than 30000ms for a time interval of 50ms and when the time constraint is less than 10000ms for a time interval of 1000ms.
\item However, when the time constraint is not stringent, such as when the time constraint is greater than 60000ms for a time interval of 50ms and when the time constraint is greater than 30000ms for a time interval of 100ms, DDS does not perform as well as EODS.
\item The overhead in the DDS algorithm exists because the scheduler on Raspberry Pi saves more images to process locally with more time, leading to longer queuing lists and many last images exceeding the time limit.
\item  in practical situations where the time interval and the time constraint are not large, DDS has the highest priority among all the scheduling algorithms for Edge AI systems.
\end{itemize}

\subsubsection{Performance of extending the end device scale  }

In certain cases, the edge server may face difficulties in processing all images received by it. The graph depicted in Figure \ref{fig:50} highlights that when there are too many images waiting to be processed on the edge server within a limited time, the processing time of the last few images tends to exceed the given time constraint. To tackle this issue, one approach is to expand the scale of end devices, allowing the main scheduler on the edge server to dynamically distribute tasks across other end devices, thereby enabling more images to meet the time constraint. Another reason to offload tasks from the edge server is to maximize the utilization of end devices' computing resources while keeping the edge server lightly loaded to effectively manage the entire system.

We set different CPU loads for the edge server to simulate its real-world scenarios where it is always occupied by many tasks. Firstly, we set an experiment to investigate the relationship between CPU load and the average processing time of containers. The experiment results showed that as the CPU load increases from 0 to 100\%, the average processing time of the containers also increases, suggesting that CPU load directly impacts container performance. At a CPU load of 0\%, the container took an average of 223 units of processing time, which increased to 284, 312, and 350 units of processing time at 25\%, 50\%, and 75\% CPU load, respectively. When the CPU load was close to 100\%, the processing time increased to 374 units, indicating that the container was under significant strain at high CPU loads.

In order to improve the performance of an edge computing system, it is not simply a matter of extending the scale of end devices. Merely increasing the number of end devices without a well-designed scheduling mechanism can lead to suboptimal performance and wasted resources. Therefore, it is crucial to prioritize the development of effective scheduling mechanisms in order to fully realize the potential benefits of edge computing.

This conference paper presents a novel contribution to the field of distributed computing: DDS with real-world performance evaluation. The proposed DDS allows end devices to update their profile, such as the number of containers running with the edge server, on a regular basis. This enables the edge server to make dynamic decisions about current tasks that meet requirements and optimize performance by processing as many tasks that meet requirements as possible simultaneously.

To ensure that the edge server is not overly burdened, the basic scheduling rule is to check whether the current task can be processed with time constraints if sent to other devices. If one end device is able to do so, the task is sent to that device. Otherwise, it is processed on the edge server. By analyzing the number of containers running and running time under specific loads from container evaluation results, the edge server calculates the predicted processing time for the current image, enabling a queuing list of tasks on the end device.

However, using a queue to store tasks introduces a time difference between decision-making and actual execution, which can reduce predicting accuracy. To mitigate this issue, the scheduler checks whether the end device has available containers, and only offloads the task to that device if containers are available. Otherwise, the edge server will process the task locally. By implementing and evaluating this DDS, we demonstrate its effectiveness in real-world scenarios and highlight its potential for large-scale deployment.

Figure \ref{fig:DDS_vs_DDSwithR2} presents a comparison between an edge server using DDS and an extended system that includes one more Raspberry Pi using the same DDS. The time interval is set to 50ms. Figure \ref{fig:5000_DDS_vs_DDSwithR2} and Figure \ref{fig:10000_DDS_vs_DDSwithR2} depict the performance under time constraints of 5000ms and 10000ms, respectively.

To evaluate the system, we stress the CPU load from 0 to 100 percent, as shown on the horizontal axis. The vertical axis indicates the number of images, out of a total of 1000, that meet the time constraint. 

These results provide valuable insights into the performance of the DDS and its extension, as well as their ability to handle varying levels of CPU load while maintaining task completion within specified time constraints. 

\begin{itemize}
\item The number of images meeting the time constraint decreases as the CPU load increases.
\item with a time constraint of 5000ms, the number of images meeting requirements decreases by approximately 20\% as the CPU load increases from 0 to 100\%, while with a time constraint of 10000ms, this decrease is about 30\%.
\item the performance of the system improves when an additional Raspberry Pi is incorporated. For instance, in Figure \ref{fig:5000_DDS_vs_DDSwithR2}, at a CPU load of 0, the number of images meeting the time constraint is 327 when using DDS alone, whereas this number increases to 551 when using DDS with an additional Raspberry Pi, indicating an improvement of 69\%.

\end{itemize}

\section{Conslusion}
% 1. run everything on the server;
% 2. network, CPU load;
% 3. time constraint, will include other types of constraints, privacy, 
% 4. now edge server + 2 rasp, will include GPU, ACICs, 

This paper proposes an integrated distributed scheduling architecture to support AI applications at the edge, which we have implemented. Additionally, we propose a new dynamic distributed scheduling approach (DDS) based on pre-evaluation of the computing capacity of all nodes in the architecture. By regularly updating the edge server with current global information, it can dynamically distribute jobs to appropriate nodes. This scheduling algorithm prioritizes two basic rules: first, to let end devices close to the data source process jobs if they are capable, and second, to take full advantage of end devices to keep the edge server's load low.

The experimental results provide evidence for the effectiveness of the DDS algorithm in comparison to other scheduling algorithms. Comparing the performance of running all tasks on the edge server versus running them on a Raspberry Pi 1 highlights the importance of a distributed scheduling mechanism. Additionally, comparing the efficiency of static versus dynamic distributed scheduling indicates the superiority of the dynamic approach.

In this paper, we focus mainly on time constraints, but we acknowledge that there are other constraints, such as privacy concerns, energy efficiency, network constraints, and others that could be considered to further improve our system. In the future, we plan to extend this architecture to include more edge servers and end devices, including powerful computing nodes such as GPUs and ASICs, to support a broader range of AI applications, such as augmented reality applications.

%\section*{References}

\end{document}